\documentclass[aps,prl,twocolumn,amsfonts,amsmath,amssymb,showpacs,final,10pt]{revtex4-1}
\usepackage[dvipsnames]{xcolor}
\usepackage[colorlinks=true,citecolor=Blue,linkcolor=Blue,urlcolor=Blue]{hyperref}
\usepackage{graphicx,manfnt,pifont,colonequals}
\usepackage{bbm}
\newcommand\noi{\noindent}

\newcommand\restr[2]{{\left.\kern-\nulldelimiterspace#1\vphantom{\big|}\right|_{#2}}}

\begin{document}

\title{\texorpdfstring{Defects, modular differential equations, and free field realization of $\mathcal{N} = 4$ VOAs}{}}
\author{Yiwen Pan, Yufan Wang, Haocong Zheng}
\affiliation{School of Physics, Sun Yat-Sen University, Guangzhou 510275, China}

\begin{abstract}
\noi For all 4d $\mathcal{N} = 4$ SYM theories with simple gauge groups $G$, we show that the residues of the integrands in the $\mathcal{N} = 4$ Schur indices, which are related to Gukov-Witten type surface defects in the theories, equal the vacuum characters of $\operatorname{rank}G$ copies of $bc\beta\gamma$ systems that provide the free field realization of associated $\mathcal{N} = 4$ VOAs in \cite{Bonetti:2018fqz}. This result predicts that these residues, as module characters, are additional solutions to the flavored modular differential equations satisfied by the original Schur index. The prediction is verified in the $G = SU(2)$ case, where an additional logarithmic solution is constructed.
\end{abstract}

\maketitle

\section{Introduction}

In \cite{Beem:2013sza}, any four-dimensional $\mathcal{N} = 2$ superconformal theory (4d $\mathcal{N} = 2$ SCFT) is shown to contain a 2d vertex operator algebra (VOA) as a protected subsector. The associated VOA encodes important information of the 4d SCFT, which can be accessed using various tools available for 2d VOAs \cite{Lemos:2015orc,Cordova:2015nma,Beem:2014rza,Lemos:2014lua}.

The correspondence immediately predicts new classes of VOAs \cite{Song:2017oew,Xie:2019yds,Xie:2019vzr,Creutzig:2017qyf} and also inspires novel realizations \cite{Bonetti:2018fqz,Beem:2019tfp,Beem:2019snk} of some known VOAs. In \cite{Bonetti:2018fqz}, the authors proposed a free field realization of the associated VOAs of the $\mathcal{N} = 4$ Super-Yang-Mills (SYM) with gauge groups $G$ in terms of $\operatorname{rank}G$ copies of $bc\beta\gamma$ systems. The construction further predicts explicitly new $\mathcal{N} = 2$ VOAs labeled by complex reflection groups.

Another particularly intriguing entry in the 4d/2d dictionary involves BPS surface defects in the 4d SCFTs. It is generally believed that they give rise to (twisted) modules of the associated 2d VOAs \cite{Cordova:2017mhb,Beem:203X}. Some important progress have been made to elucidate this relation \cite{Cordova:2017mhb,Cordova:2016uwk,Nishinaka:2018zwq,Pan:2017zie,Dedushenko:2019yiw}, where a UV class-$\mathcal{S}$ theory $\mathcal{T}'$ is Higgssed, with a position-dependent vev \cite{Gaiotto:2012xa}, to an IR theory $\mathcal{T}$ coupled to certain surface defect. At the level of Schur indices, the UV Schur index is related to the IR index with the surface defect by a simple formula,
\begin{align}
  \mathcal{I}_\text{IR + defect} = \lim\mathcal{I}_\text{NG}^{-1}\mathcal{I}_\text{UV}\ ,
\end{align}
where the limit is taken for suitable combination of flavor fugacities that encodes the position-dependence of the operator getting a vev during Higgsing. The $\mathcal{I}_\text{UV}$ and $\mathcal{I}_\text{IR + defect}$ denote respectively the Schur index of the theory before Higgsing, and that of the resulting IR theory with a surface defect after Higgsing. The $\mathcal{I}_\text{NG}^{-1}$ removes divergences associated to the decoupled Nambu-Goldstone multiplets.

One simple class of surface defects attracts relatively less attention in the literature on this particular 4d/2d correspondence, namely, those engineered by a singular BPS background profile of a dynamical gauge field in a 4d $\mathcal{N} = 2$ SCFT \cite{Gukov:2008sn}. In this paper, we will consider these defects in 4d $\mathcal{N} = 4$ SYM with simple gauge groups $G$. Their Schur indices with the surface defects are computed by shifting the integration variables in the one-loop determinant \cite{Drukker:2012sr,Kapustin:2012iw,Nawata:2014nca,Hosomichi:2017dbc} in the original Schur index, and therefore related to the residues of the integrands. What is surprising is that these residues precisely coincide with the vacuum characters of the $\operatorname{rank}G$ copies of $bc\beta \gamma$ systems responsible for the free field realization \cite{Bonetti:2018fqz}. This observation, first made in \footnote{W. Peelaers, private communication on the cases with $\mathcal{N} = 4$ $SU(2)$ and $SU(2)$ SQCD theories.}, facilitates the identification of the modules of the associated $\mathcal{N} = 4$ VOAs that correspond to the surface defects, such that the Schur indices with or without surface defects can be written as combinations of module characters. We study in detail the simplest example of 4d $\mathcal{N} = 4$ theory with the gauge group $SU(2)$, where we identify the combinations of simple modules corresponding to the surface defect. Vanishing one-point function of null vectors of the VOA evaluated on modules sometimes lead to flavored modular differential equations (FMDEs) satisfied by the module characters. To verify this prediction we check that the Schur indices with/without defect satisfy the FMDEs predicted by the nulls studied in \cite{Beem:2017ooy}, and we further construct an
additional logarithmic solution using the module characters.

Computationally, our results highlight a fact that the simple \emph{integrand} of a Schur index (instead of the full index as a contour integral) of a Lagrangian theory directly encodes crucial, yet easily accessible, structural information of the associated VOA.


\section{Review of localization}

The flavored Schur index $\mathcal{I}$ of a 4d $\mathcal{N} = 2$ SCFT $\mathcal{T}$ is defined as a supertrace ($\operatorname{str}$) over the Hilbert space,
\begin{align}
  \mathcal{I} = q^{c_\text{4d}/2}\operatorname{str} q^{E - R} \mathbf{b}^\mathbf{f} \ ,
\end{align}
where $c_\text{4d}$ is the $c$-central charge of $\mathcal{T}$, $E$ is the conformal dimension, $R$ is the $SU(2)_\mathcal{R}$ charge, $\mathbf{b}$ and $\mathbf{f}$ collectively denote flavor fugacities and flavor Cartan generators. We will parametrize $q = e^{2\pi i \tau}$ with $|q|< 1$. The index coincides with the vacuum character of the associated VOA $\mathcal{V}_{\mathcal{T}}$ \cite{Beem:2013sza}. Furthermore, for Lagrangian theories the index and the flavored character respectively can be identified with $S^3 \times S^1_t$ and $T^2 = S^1 \times S_t^1$-partition functions.

This chain of identifications is clarified in \cite{Pan:2019bor,Dedushenko:2019yiw} using supersymmetric localization. Geometrically, the $S^3$ is traditionally parametrized by $\varphi, \chi \in [0, 2\pi]$ and $\theta \in [0, \frac{\pi}{2}]$, where $\theta = 0, \pi/2$ represent two separate and linked tori $T^2$ and $T^2_{\theta = \frac{\pi}{2}}$. Note that in the background for localization, the torus $T^2$ has complex moduli given by $\tau$. On the BPS locus, the vector multiplet gauge field is flat with other component fields set to zero. Smoothness and flatness requires that $A_\text{BPS} = \mathfrak{a}dt$ where $\mathfrak{a}$ is a constant Cartan-valued element. The hypermultiplet scalars are covariantly constant along $\chi$ and constrained by elliptic partial differential BPS equations in the $t, \varphi, \theta$ directions, and therefore their BPS configurations are determined by the boundary values $\beta, \gamma$ on $T^2$.

In the end, the Schur index localizes to some free $bc \beta \gamma$ system on the $T^2$, and can be written explicitly as \footnote{Here the prime in the measure indicates that we drop the $c$ zero mode when it is $\mathfrak{su}(2)$ Cartan-valued.}
\begin{align}
  \mathcal{I} = \oint_{|a| = 1} \left[\frac{da}{2\pi i a}\right] \int [DbDc]'[D\beta D\gamma] e^{- S_{bc \beta \gamma}[\mathfrak{a}, \mathfrak{b}]} \ ,
\end{align}
where $[da/2\pi i a]$ the Haar measure for integration,
\begin{align}
  S_{bc\beta\gamma}[\mathfrak{a}, \mathfrak{b}] = \int_{T^2} (\beta D_{\bar z} \gamma + b D_{\bar z}c) \ ,
\end{align}
and in the covariant derivative $D_{\bar z} = \partial_{\bar z} - i \rho(A_{\bar z}) - i \rho(A^\text{flavor}_{\bar z}) = \partial_{\bar z} + i \frac{\rho(\mathfrak{a} + \mathfrak{b})}{\tau - \bar \tau}$ which acts on a field in the gauge/flavor representation $\rho$, we have also included a flavor background gauge field $A^\text{flavor} = \mathfrak{b}dt$. We will frequently use the exponentiated fugacities $q = e^{2\pi i \tau}$, $a = e^{2\pi i \mathfrak{a}}$, $b = e^{2\pi i \mathfrak{b}}$.

The localization computation can be enriched by Schur operator insertions. In \cite{Pan:2019bor}, these operators are shown to be almost BPS and ``localizable'' by a generalized localization argument. Given that these Schur operators are gauge-invariant combinations of the free $bc \beta \gamma$ fields, their correlation functions can be computed by inserting free field torus correlators (obtained from suitable $bc \beta \gamma$ Green's functions and the standard Wick's theorem) into the contour integral.

\section{\texorpdfstring{Defects and free field characters}{}}

Let us consider a 4d $\mathcal{N} = 4$ SYM with a simply-connected simple gauge group $G$ with Lie algebra $\mathfrak{g}$. The Schur index is well-known, given by
\begin{align}
  \mathcal{I}(a, b) \equiv \frac{{(-1)^{|\Delta^+|}}}{|W|}\oint & \ \left[\frac{da}{2\pi i a}\right]
  \frac{\eta(\tau)^{3r}}{\vartheta_4(\mathfrak{b}|\tau)^{r}} \nonumber\\
  & \ \times \prod_{\alpha \in \Delta} \frac{\vartheta_1(\alpha(\mathfrak{a})|\tau)}{\vartheta_4(\alpha(\mathfrak{a}) + \mathfrak{b}|\tau)} \ .
\end{align}
Here $\mathfrak{b}$ (with $b \equiv e^{2\pi i \mathfrak{b}}$) is the flavor symmetry fugacity, $\mathfrak{a} \in \mathfrak{h}$ (also with $a \equiv e^{2\pi i \mathfrak{a}}$) denotes the Cartan-valued flat connection along the temporal $S^1$ and $[da/2\pi i a]$ an appropriate measure \footnote{When writing the integrand in terms of $q$-Pochhammer symbols, some polynomial factors from the $\vartheta_1$'s will join $[da/2\pi i a]$ to form the standard Haar measure.}. Also, $r$ denotes the rank of $\mathfrak{g}$, $\Delta$ ($\Delta^\pm$) the set of all roots (positive/negative roots) and $|\Delta|$ the number of roots. As usual $\alpha_{i = 1, \ldots, r}$ denotes the simple roots of $\mathfrak{g}$.

It will prove convenient to decompose all positive roots according to their ``height'' $H$ (if $\alpha = \sum_{i = 1}^{r}m_i \alpha_i$, then $H(\alpha) \equiv \sum_{i = 1}^{r} m_i$), and use $\Delta^+_H$ to collect all positive roots of height $H$. With such definition, we rewrite all products over roots into
\begin{align}
  \prod_{\alpha \in \Delta} f(\alpha) = \prod_{H \ge 1} \prod_{\alpha \in \Delta^+_H}f(\alpha) \prod_{\alpha \in \Delta^+_H} f( - \alpha) \ .
\end{align}

We are interested in the theory in the presence of a Gukov-Witten type surface defect specified by a background gauge field with a profile $A^\text{bg} = \mathfrak{a}_\varphi d\varphi $ where $\mathfrak{a}_\varphi \in \mathfrak{h} $, which is singular on the torus $T^2_{\theta = \pi/2}$. Other component fields in the vector multiplet are set to zero. This background configuration is BPS with respect to the supercharge used for localization. As a result, $A^\text{bg}$ modifies the final path integral on $T^2$ that computes the Schur index \footnote{A complete treatment involves first removing a tubular neighborhood of the surface defect and consider suitable fields and supersymmetric theory on that space, along the line as \cite{Drukker:2012sr,Hosomichi:2017dbc}. The neighborhood does not intersect the $T^2$ where the $bc \beta \gamma$ system lives, and we conjecture that the only effect left on $T^2$ is the background gauge field $\mathfrak{a}_\varphi$. We leave the precise computation to future work.}, which now reads,
\begin{align}
  \label{twisted-path-integral}
   \oint \frac{da}{2\pi i a} \int [DbDc]'[D\beta D\gamma] e^{- S_{bc \beta \gamma}[\mathfrak{a}, \mathfrak{a}_\varphi, \mathfrak{b}]} \ .
\end{align}
The torus action here is simply
\begin{align}
  \label{twisted-action}
  S_{bc\beta\gamma}[\mathfrak{a}, \mathfrak{a}_\varphi, \mathfrak{b}] = \int_{T^2} (\beta D_{\bar z} \gamma + b D_{\bar z}c) \ ,
\end{align}
where $D_{\bar z} = \partial_{\bar z} - i A_{\bar z} - i A^\text{bg}_{\bar z} - i A^\text{flavor}_{\bar z} = \partial_{\bar z} + i \frac{(\mathfrak{a} - \tau \mathfrak{a}_\varphi + \mathfrak{b})}{\tau - \bar \tau}$. When $\mathfrak{a}_\varphi = 0$, the path integral recovers the original index. For non-zero $\mathfrak{a}_\varphi$ the $bc \beta \gamma$ systems are taken to the twisted sector labeled by $\mathfrak{a}_\varphi$, whose character can be computed \cite{Eholzer:1997se}. In the end, the index in the presence of the surface defect reads
\begin{align}
  \mathcal{I}^\text{defect}(a, b) \equiv \frac{{(-1)^{|\Delta^+|}}}{|W|}& \ \oint \left[\frac{da}{2\pi i a}\right]
  \frac{\eta(\tau)^{3r}}{\vartheta_4(\mathfrak{b}|\tau)^{r}}\\
  & \ \times \prod_{\alpha \in \Delta} \frac{\vartheta_1(\alpha(\mathfrak{a} - \tau \mathfrak{a}_\varphi)|\tau)}{\vartheta_4(\alpha(\mathfrak{a} - \tau \mathfrak{a}_\varphi) + \mathfrak{b}|\tau)} \ . \nonumber
\end{align}
We can absorb the shift by $\mathfrak{a}_\varphi$ into the integration variables which effectively shift their contours from unit circles $|a_i| = 1$ to $|a_i| = |q^{- (\mathfrak{a}_\varphi)_i}|$, where we define $\mathfrak{a}_\varphi = \operatorname{diag}( (\mathfrak{a}_\varphi)_{i = 1, \ldots, N}) \in \mathfrak{h}$.

We argue that this defect index is related to the residues of the integrand. For simplicity we take $G = SU(3)$ temporarily and give $\mathfrak{b}$ a small positive imaginary part so that $|b| < 1$. Now, imagine we gradually turn on the defect parameter $(\mathfrak{a}_\varphi)_1$ from 0 to $- \frac{1}{2}$, which shrinks the $a_1$ contour from $|a_1| = 1$ to $|a_1| = |q|^{\frac{1}{2}}$. The integral stays the same initially, but right before reaching the final contour, the pole $a_1 = a_1^* \equiv a_2 b^{-1}q^{\frac{1}{2}}$ from the factor $\vartheta_4(\mathfrak{a}_1 - \mathfrak{a}_2 + \mathfrak{b})^{-1}$ crosses the shrunk $a_1$ contour. As $a_1$-contour shrinks, the $a_2$-pole $a_2^2 = (a^*_2)^2 \equiv a_1^{-1} b^{-1} q^{\frac{1}{2}}$ from the factor $\vartheta_4(\mathfrak{a}_1 - \mathfrak{a}_3 + \mathfrak{b})^{-1}$ actually starts venturing outward from inside the $a_2$-unit circle. In the end, the pole reaches $|a_2^2| = |b^{-1}|$ which is outside of the $a_2$-integration contour: the pole $a_2^*$ crosses the $a_2$-contour precisely when the pole $a_1^*$ crosses the $a_1$-contour \footnote{Note that another potential pole $(a_1 = a_2 b^{-1}q^{\frac{1}{2}}, a_2 = a_1^{- 2} b^{-1}q^{\frac{1}{2}})$ is canceled by the $\vartheta_1(\mathfrak{a}_2 - \mathfrak{a}_3)^2$ factor.}. At the end of the movement, $\alpha_i(\mathfrak{a}_\varphi) = - \frac{1}{2}$ for both $i = 1, 2$, and the defect index equals the original Schur index $\mathcal{I}$ with the residue $\operatorname{Res}$ of the simultaneous pole $(\frac{a_1}{a_2} = b^{-1}q^{\frac{1}{2}}, \frac{a_2}{a_3} = b^{-1}q^{\frac{1}{2}})$ discussed above subtracted,
\begin{align}
  \mathcal{I}^\text{defect} = \mathcal{I} - \operatorname{Res}\ .
\end{align}
Moreover, by direct computation, other poles that one might encounter by different ways of turning on $\mathfrak{a}_\varphi$ actually share the same residue, up to numeric constants and a power of $q$, as analytic function of $\mathfrak{b}$ and $q$ \footnote{Although equivalent as analytic functions, these residues are different as $\mathfrak{b}$-series with specific convergence annulus. We hope to come back to this issue in future work.}. One should carefully collect all the poles that crosses the contour when gradually turning on the defect. In the following discussion for more general simple gauge groups $G$, we will focus on the simplest set of simultaneous poles, and leave the full discussion on other poles to future study.

Let us now consider the simple algebra $\mathfrak{g}$ of a simple Lie group $G$, and focus on the (simultaneous) poles from the $\vartheta_4$'s in the denominator given by the equations \cite{Nishinaka:2018zwq,Goldstein:2020yvj}
\begin{align}
  e^{ 2\pi i \alpha_i (\mathfrak{a})} = bq^{\frac{1}{2}} \ , \qquad i = 1, \ldots, r \ .
\end{align}
These equations imply that $e^{2\pi i \alpha(\mathfrak{a})} = (bq^{\frac{1}{2}})^H$ for $\forall \alpha \in \Delta_H^+$. 

It is straightforward to compute the residue of the full integrand at the poles \cite{Nishinaka:2018zwq,Goldstein:2020yvj}. We first present the raw result before further massage (we have dropped the overall sign and $1/|W|$ to avoid clutter, and written the theta functions in terms of $(z;q)$), which reads
\begin{align}
  \operatorname{Res} = & \ \frac{q^{\frac{|\Delta| + r}{8} }(q;q)^{3r}}{[(q;q)(bq^{\frac{1}{2}};q)(b^{-1}q^{\frac{1}{2}};q)]^{r}} \frac{1}{(q;q)^r}\nonumber\\
  \times & \ \prod_{H \ge 1}\frac{[( (bq^{\frac{1}{2}})^Hq;q )((bq^{\frac{1}{2}})^{ - H};q)]^{|\Delta^+_H|}}{[( (bq^{\frac{1}{2}})^{H + 1};q)( (bq^{\frac{1}{2}})^{- H - 1} q;q)]^{|\Delta^+_H|}} \\
  \times & \ \prod_{H \ge 1} \frac{
    [( (bq^{\frac{1}{2}})^{-H} q;q )( (bq^{\frac{1}{2}})^{H} ;q )]^{|\Delta_H^+|}
  }{
    [( (bq^{\frac{1}{2}})^{-H + 1} ;q )'( (bq^{\frac{1}{2}})^{H - 1} q;q )]^{|\Delta_H^+|}
  } \ .\nonumber
\end{align}
Here the prime in the last line indicates that the corresponding factors with $H = 1$ are dropped; they are accounted for by the $(q;q)^{-r}$ in the first line, where the $q$-Pochhammer symbol is defined by $(z;q) \equiv \prod_{k = 0} (1 - zq^k)$.

It is obvious that there are massive cancellation between the second and the third line. Concretely, \emph{almost} every factor of $(\#;q)^\#$ (say, corresponding to a height $H$) from the second line will find its opponent (corresponding to $H + 1$) in the third line:
\begin{itemize}
  \item If a height $H$ is such that $|\Delta_H^+| = |\Delta^+_{H + 1}|$, then the two factors completely annihilate each other. However, if $|\Delta_H^+| > |\Delta_{H+1}^+|$, part of the factor from the second line survives.
  \item Additionally, the factors in the second line with the the largest $H$ will find no match and therefore always survive.
  \item The factors in the third line with $H = 1$ will cancel against those in the first line. Note that $|\Delta_{H = 1}^+| = r$.
\end{itemize}
Finally, with all these cancellations carried out, we are left with
\begin{align}
  q^{\frac{\dim \mathfrak{g}}{8}}\prod_{\substack{H \ge 1 \\  |\Delta_H^+| > |\Delta_{H + 1}^+|}} \frac{(b^H q^{\frac{1}{2} + \frac{H + 1}{2}};q) (b^{- H} q^{\frac{1}{2} - \frac{H + 1}{2}};q)}{(b^{H + 1} q^{\frac{H + 1}{2}};q) (b^{ - (H + 1)} q^{1 - \frac{H + 1}{2}};q)} \ .
\end{align}
\begin{table}
  \begin{tabular}{c|c|c}
    $\mathfrak{g}$ & $H$ & $|\Delta_H^+|$\\
    \hline
    $\mathfrak{a}_r$ & $1, 2, ..., r$  & $r, r - 1, \ldots, 1$\\
    $\mathfrak{d}_{2n}$ & $1, 2, ..., 2n - 1$  & $2n, (2n-1)^2, (2n -2)^2, \ldots, (n + 1)^2 $\\
    & $\qquad 2n, \ldots, 4n - 3$ & $ \qquad \qquad (n-1)^2  \ldots , 1^2$\\
    $\mathfrak{e}_6$ & $1, 2, \ldots, 11$ & $6, 5^3, 4, 3^2, 2, 1^3$\\
    $\mathfrak{e}_7$ & $1, 2, \ldots, 17$ & $7,6^4, 5^2, 4^2, 3^2, 2^2, 1^4$\\
    $\mathfrak{e}_8$ & $1, 2, \ldots, 29$ & $8, 7^6,6^4,5^2, 4^4, 3^2, 2^4, 1^6$\\
  \end{tabular}
  \caption{The number of roots at each $H$ for some simplest simple Lie algebras. In the third column, $n_1^{m_1}, n_2^{m_2}, \ldots$ encodes that at $m_i$ \emph{consecutive} heights there are $n_i$ roots inside. For example, in the row of $\mathfrak{e}_6$, $6,5^3, 4, \ldots$ means that there are $6, 5,5,5,4, \ldots$ roots at height $1, 2, 3, 4, 5, \ldots$. From these data we can read off at which height the number of roots decreases.\label{heights}
  } 
\end{table}

\begin{table}
  \begin{tabular}{c|c}
    $\mathfrak{g}$ & $d_1, \ldots, d_r$\\
    \hline
    $\mathfrak{a}_r$ & $1, 2, \ldots, r$ \\
    $\mathfrak{b}_r$ & $2, 4, \ldots, 2r$ \\
    $\mathfrak{d}_{n}$ & $2, 4, \ldots, 2(r - 1); r$\\
    $\mathfrak{e}_6$ & $2,5,6,8,9,12$ \\
    $\mathfrak{e}_7$ & $2,6,8,10,12,14,18$\\
    $\mathfrak{e}_8$ & $2,8,12,14,18,20,24,30$\\
    $F_4$ & $2,6,8,12$\\
  \end{tabular}
  \caption{The degrees of invariants of the simple Lie algebras.\label{degrees}} 
\end{table}

Here we have used the fact that for all simple Lie algebras, $0 \le |\Delta_H^+| - |\Delta_{H + 1}^+|$. In fact, when the inequality is a strict inequality, $H + 1$ coincides with the degree of an invariant of $\mathfrak{g}$ \cite{Collingwood} \footnote{note the special multiplicity of $2$ for the $\mathfrak{d}_r$ case at $H = r - 1$, since $|\Delta_{r - 1}^+| - |\Delta_r^+| = 2$, so the corresponding factor in the above product should be squared.}, which further agrees with the degree of a fundamental invariant of the associated Weyl group! See Table \ref{heights} and Table \ref{degrees} for concrete examples. Hence, we finally recognize the residue to be
\begin{align}
  q^{\frac{\dim\mathfrak{g}}{8}}\prod_{i = 1}^r \frac{(b^{d_i - 1} q^{ \frac{d_i + 1}{2}};q) (b^{- d_i + 1} q^{\frac{1 - d_i}{2}};q)}{(b^{d_i} q^{\frac{d_i}{2}};q) (b^{ - d_i} q^{1 - \frac{d_i}{2}};q)} \ ,
\end{align}
which is precisely the vacuum character of the $\operatorname{rank}\mathfrak{g}$ copies of $bc\beta \gamma$ systems appearing in the free field realization of $\mathcal{N} = 4$ VOAs \cite{Bonetti:2018fqz}, namely (up to some numerical constants)
\begin{align}
  \operatorname{Res} = \operatorname{ch}(V^G_{bc \beta \gamma})\ .
\end{align}
In particular, the central charge matches as expected, $c = -3 \dim \mathfrak{g} = -3 \sum_{i = 1}^{r} c_{bc\beta \gamma}^i$.


As proposed in \cite{Bonetti:2018fqz}, the $\mathcal{N} = 4$ VOA $\mathcal{V}^{G}_{\mathcal{N}=4}$ is embedded as a subalgebra in the $bc \beta \gamma$ system $\mathcal{V}^{G}_{bc\beta \gamma}$. Consequently, the vacuum module $V^{G}_{bc\beta \gamma}$ also furnishes a reducible but indecomposable module of the VOA $\mathcal{V}^{G}_{\mathcal{N}=4}$, with the vacuum module $V^G_{\mathcal{N} = 4}$ of $\mathcal{V}^G_{\mathcal{N} = 4}$ a submodule of $V^{G}_{bc\beta\gamma}$.

If $N \in \mathcal{V}^G_{\mathcal{N} = 4}$ is a null vector, then one can insert $N$ into the supertrace over any module $M$ of $\mathcal{V}_{\mathcal{N} = 4}^G$ and the result should vanish. As discussed in \cite{Beem:2017ooy,Gaberdiel:2008pr,Gaberdiel:2008ma,Beem:202X}, for $N$ of a special type, one can derive from the supertrace a (flavored) modular differential equation,
\begin{align}
  0 = \operatorname{str}_M N(z) q^{L_0 - \frac{c_\text{2d}}{24}} b^f = \mathcal{D}_{q,b} \operatorname{ch}(M) \ ,
\end{align}
where $\mathcal{D}_{q,b}$ denotes a differential operator with simple modular property. In particular, by choosing $M = V^G_{bc \beta \gamma}$ one concludes that the residue $\operatorname{Res}$ must be a solution to all the flavored modular differential equations predicted by the nulls of $\mathcal{V}^G_{\mathcal{N} =4}$. Next we will elaborate on the simplest case with $G = SU(2)$.

\section{\texorpdfstring{Example: 4d $\mathcal{N} = 4$ $SU(2)$-SYM}{}}

The associated VOA of the 4d $\mathcal{N} = 4$ Super-Yang-Mills with an $SU(2)$ gauge group is the small $\mathcal{N} = 4$ superconformal algebra $\mathcal{V}_{\mathcal{N} = 4}$ with $c_\text{2d} = -9$, generated by $J^a, G^\pm, \tilde G^\pm$ \cite{Beem:2013sza}. Here $\{J^a\}$ generate an $\widehat{\mathfrak{su}}(2)_{k = - \frac{3}{2}}$ affine subalgebra, and at this central charge the Sugawara stress tensor coincides with that of the full VOA $\mathcal{V}_{\mathcal{N} = 4}$. The flavored Schur index of the theory, or equivalently, the flavored vacuum character of the VOA can be written as the standard contour integral,
\begin{align}\label{Schur-index}
  \mathcal{I} = - \frac{1}{2} \oint_{|a| = 1} \frac{da}{2\pi i a} \vartheta_1(\pm 2\mathfrak{a}) & \ \prod_{n = -1}^{+1}\frac{\eta(\tau)}{\vartheta_4(2n \mathfrak{a} )} \nonumber \\
  \equiv & \ \oint \frac{da}{2\pi i z} Z(a, b;q) \ .
\end{align}

As explained in the previous section, we are interested in the theory with a surface defect engineered by turning on a background gauge field of the form $A^\text{bg} = \mathfrak{a}_\varphi  \operatorname{diag}(1, -1) d\varphi$. Here $\mathfrak{a}_\varphi \operatorname{diag}(1, -1) \in \mathfrak{h} \subset \mathfrak{su}(2)$. The defect index is then written as

\begin{align}
  \mathcal{I}^\text{defect}(b, q) = \oint_{|a| = |q^{- \mathfrak{a}_\varphi}|} \frac{da}{2\pi i a}Z(a, b; q) \ ,
\end{align}
where we have absorbed the $\mathfrak{a}_\varphi$ into the integration variable which deforms the contour accordingly. The index equals the original Schur index if $|\mathfrak{a}_\varphi|$ is relatively small since the integrand is meromorphic in $a$. However, as $\mathfrak{a}_\varphi$ varies from $0$ towards larger negative values, say, $\mathfrak{a}_\varphi = - \frac{1}{4}$, the shrunk contour will inevitably hit the poles corresponding to $ 2 \mathfrak{a} + \mathfrak{b} = \frac{\tau}{2}$. Concretely, the poles are given by $a = \pm b^{- \frac{1}{2}}q^{\frac{1}{4}}$, whose total residue will be denoted as $\operatorname{Res}(b, q)$. As a result, the defect index reads
\begin{align}
  \mathcal{I}^\text{defect}(b;q) = \mathcal{I}(b,q) - \operatorname{Res}(b,q) \ ,
\end{align}
where the residue equals to
\begin{align}
  \operatorname{Res}(b,q) \equiv & \ - \frac{1}{2} q^{\frac{3}{8}} \frac{(b^{-1}q^{-\frac{1}{2}};q)(b q^{\frac{3}{2}};q)}{(b^{-2};q)(b^2q;q)} \nonumber \\
  = & \ - \frac{1}{2} \frac{1}{1 - b^{ - 2}} \Big( - bq^{\frac{1}{8}}
    + (1 + b^{ - 2})q^{\frac{3}{8}} \\ 
    &\qquad\qquad - (b^{+1} + b^{-1} + b^{ - 3})q^{\frac{7}{8}} + \ldots \Big)\nonumber
\end{align}
Note that the residue is singular in the $b \to 1$ limit, and hence $\operatorname{Res}$ and $\mathcal{I}$ are linear independent. 

As observed in the previous section, the above residue, up to the factor $- \frac{1}{2}$, is nothing but the character of the character of the vacuum module $V_{bc\beta\gamma}$ of a $bc \beta \gamma$ system $\mathcal{V}_{bc\beta \gamma}$ studied in \cite{Bonetti:2018fqz}, which is responsible for the free field realization of the small $\mathcal{N} = 4$ superconformal algebra $\mathcal{V}_{\mathcal{N} = 4}$,
\begin{align}
  - 2\operatorname{Res}(b,q) = \operatorname{ch}(V_{bc\beta \gamma}) \equiv \operatorname{str}_{V_{bc\beta \gamma}} q^{L_0 - \frac{c_\text{2d}}{24}} b^f\ .
\end{align}
For convenience, we reproduce here their conformal weights and the $U(1)$-charge,
\begin{center}
  \begin{tabular} {c|c|c}
    & $h$ & $m$\\
    \hline
    $(b, c)$& $(\frac{3}{2}, - \frac{1}{2})$ & $(\frac{1}{2}, - \frac{1}{2})$ \\
    $(\beta, \gamma)$& $(1,0)$& $(1, -1)$ \\
  \end{tabular}
\end{center}
Some generators of the small $\mathcal{N} = 4$ superconformal algebra $\mathcal{V}_{\mathcal{N} = 4}$ written in the free field realization are
\begin{align}
  J^+ = \beta , \quad G^+ = b\ , \quad T = -\frac{3}{2}b\partial c - \frac{1}{2} \partial b c - \beta \partial \gamma \ .
\end{align}

Given that $\mathcal{V}_{\mathcal{N} = 4}$ is a sub-VOA of the $bc\beta \gamma$ system $\mathcal{V}_{bc\beta \gamma}$, its vacuum module $V_{bc\beta \gamma}$ furnishes a reducible but indecomposable module of the $\mathcal{N} = 4$ VOA $\mathcal{V}_{\mathcal{N} = 4}$. This fact immediately predicts that the residue $\operatorname{Res}$ and therefore the defect index $\mathcal{I}^\text{defect}$ must satisfy all the relevant flavor modular differential equations coming from the nulls in the $\mathcal{N} = 4$ VOA.

A simplest null worth studying is $\mathcal{N} = T - T_\text{Sug} = 0$, since the stress tensor of the $\mathcal{N} = 4$ VOA coincides with the Sugawara stress tensor of the $\widehat{\mathfrak{su}}(2)_{k = - \frac{3}{2}}$ affine subalgebra. One can directly compute the vacuum expectation value $\langle T - T_\text{Sug}\rangle$ by localization. On the one hand, the stress tensor $T$ of the 2d VOA descends from the $SU(2)_\mathcal{R}$ current in the original four dimensional theory, while on the other hand, the Sugawara stress tensor $T_\text{Sug} = \frac{1}{2(k + h^\vee)} \sum_{a,b}J^a J^b$ where the currents $J^a$ are gauge-invariant bilinears of $\beta \gamma$, $J^{+, 3, -} \sim \operatorname{tr}(\beta \beta), \operatorname{tr}(\beta \gamma)$, $\operatorname{tr}(\gamma \gamma)$. Using the $\mathfrak{a}_\varphi$-twisted Green's functions (\ref{bc-greens}, \ref{beta-gamma-greens}) for the $bc$ and $\beta \gamma$ systems and Wick theorem, we have
\begin{align}
  \langle T\rangle = \oint_{|a| = |q^{- \mathfrak{a}_\varphi}|} \frac{da}{2\pi i a}\frac{1}{4\pi^2}\left[ \frac{\vartheta_1''(2\mathfrak{a})}{\vartheta_1(2\mathfrak{a})}
  + \frac{\vartheta_1'''(0)}{2\vartheta_1'(0)}
  + \ldots
  \right]\ ,\nonumber
\end{align}
and
\begin{align}
  \langle T_\text{Sug}\rangle = \oint_{|a| = |q^{- \mathfrak{a}_\varphi}|} \frac{da}{2\pi i a}\frac{3}{8\pi^2} \left[ \frac{\vartheta''_1(0)^2}{\vartheta_1'(0)^2} + \frac{\vartheta_1'''(0)}{\vartheta_1(0)}
  + \ldots
  \right] \ . \nonumber
\end{align}
To avoid clutter we only display the first few terms. These integral at small $|\mathfrak{a}_\varphi|$ can be evaluated by picking up the residue at the origin, while for $\mathfrak{a}_\varphi = - \frac{1}{4}$, the residues at $a = \pm b^{ - \frac{1}{2}}q^{\frac{1}{4}}$ need to be subtracted off. In either case, we find $\langle T - T_\text{Sug}\rangle = 0$.

The null equation $\langle T - T_\text{Sug}\rangle = 0$ for general $\mathfrak{a}_\varphi$ is in fact a particular \emph{flavored modular differential equation} \cite{Gaberdiel:2008pr,Gaberdiel:2008ma,Arakawa:2016hkg,Beem:2017ooy,Beem:202X}. Indeed, it is straightforward to show that the integral equation can be massaged into \footnote{With help from Mathematica to establish some integral identities between integrated $q$-series. It would be great to actually prove the equality analytically.}
\begin{align}\label{fMDE1}
  q \frac{\partial}{\partial q}I = \frac{1}{2(k+h^\vee)}& \Big(\frac{1}{2}D_b^2 + k E_2 \\
  & \ + 2k  \ E_2\left[ \begin{matrix}
    1\\b^2
  \end{matrix} \right]  + 2E_1\left[\begin{matrix}
    1 \\ b^2
  \end{matrix}\right]D_b
  \Big)I\ , \nonumber
\end{align}
for $I = \mathcal{I}, \mathcal{I}^\text{defect}$, and $D_b \equiv b \partial_b$; see appendix \ref{app:FMDE} for more detail. In fact, one can further check that both indices also satisfy the flavored modular differential equations corresponding to the nulls studied in \cite{Beem:2017ooy}. See also \cite{Beem:202X} for more on FMDEs and their solutions. For example, the nulls in the eq (6.19) and eq (6.21) in \cite{Beem:2017ooy} lead to
\begin{align}
  0 = q\frac{\partial}{\partial q}& \left(b\frac{\partial }{\partial b}\right)I \nonumber\\
  & + E_1\left[
  \begin{matrix}
    -1\\b
  \end{matrix}
  \right] q\frac{\partial}{\partial q}I - 3E_3\left[
  \begin{matrix}
    -1\\b
  \end{matrix}
  \right]I
  + 6E_3\left[
  \begin{matrix}
    1\\b^2
  \end{matrix}
  \right]I\nonumber\\
  & + \left(E_2 + E_2\left[
  \begin{matrix}
    -1\\b
  \end{matrix}
  \right]
  - 2 E_2 \left[
  \begin{matrix}
    1\\b^2
  \end{matrix}
  \right]
  \right) b \frac{\partial}{\partial b}I \ ,
\end{align}
and
\begin{align}
  & \ (D_q^{(2)} + \frac{c_\text{2d}}{2} E_4 )I \nonumber \\
  & \ + \left(-2 E_2\left[
    \begin{matrix}
      -1\\b
    \end{matrix}
    \right]D_q^{(1)}
    - 4E_3\left[
    \begin{matrix}
      -1 \\b
    \end{matrix}
    \right]b \partial_b
    + 18 E_4\left[
    \begin{matrix}
      -1 \\ b
    \end{matrix}
    \right]
  \right) I \nonumber\\
  & + \left(
  3k_\text{2d}E_4 + 2 E_3\left[
  \begin{matrix}
    1 \\ b^2
  \end{matrix}
  \right] b \partial_b
  - 9 E_4\left[
  \begin{matrix}
    1 \\ b^2
  \end{matrix}
  \right]
  \right) I = 0 . 
\end{align}

Finally, let us define the quotient module $M \equiv V_{bc\beta \gamma}/V_{\mathcal{N} = 4}$. It is shown in \cite{Adamovic:2014lra} that $M$ and the vacuum module $V_{\mathcal{N} = 4}$ are the only two irreducible $\mathcal{V}_{\mathcal{N} = 4}$-modules from the category $\mathcal{O}$. Note that the space $M(0) \subset M $ of lowest conformal weight (\emph{i.e.}, $-\frac{1}{2}$) is an infinite-dimensional irreducible $\mathfrak{su}(2)$-representation with highest weight $- \omega_1$, and is spanned by $\{\gamma_0^nc_{\frac{1}{2}}\mathbf{1}| n \in \mathbb{N}\}$. The zero mode $\gamma_0$ lowers the $U(1)$ charge $2m$ by 2, which explains the factor $(1 - b^{-2})^{-1}$ in front of $\operatorname{Res}(b,q)$. As such, we identify
\begin{align}\label{defect-index}
  \mathcal{I}^\text{defect} = \frac{3}{2} \mathcal{I} + \frac{1}{2} \operatorname{ch}(M)\ ,
\end{align}
where $\operatorname{ch}(M) = - 2 \operatorname{Res} - \mathcal{I}$ is the quotient character, which by construction is a solution to all the flavored modular differential equations mentioned above. Furthermore, one can check that
\begin{align}
  \label{log-solution}
  \log b \operatorname{ch}(M) + (\log q + \log b)\mathcal{I}
\end{align}
is actually an additional logarithmic solution to all the modular differential equations, which is due to the fact that (\ref{log-solution}) arises as a modular transformation of the Schur index $\mathcal{I}$ \cite{YP:202X}.

\section{Discussions}

In this paper we study a Gukov-Witten type surface defects with prescribed singularity for the gauge fields in 4d $\mathcal{N} = 4$ SYMs with simple gauge groups from their Schur indices, which are determined by the residues of the integrands of the contour integrals that compute the original Schur indices. These residues coincide precisely with the vacuum characters of the $bc\beta\gamma$ systems in a free field realization of the $\mathcal{N} = 4$ VOAs. This observation leads to new and easily accessible solutions to the flavored modular differential equations associated to some nulls in the VOAs.

In the original construction \cite{Beem:2013sza}, the $\mathcal{N} = 4$ VOA of a $\mathcal{N} = 4$ theory is obtained via a BRST reduction of $\dim G$ copies of $bc\beta \gamma$ systems. The above computation reveals a half-way passage from that free field theory to $\operatorname{rank}G$ copies of $bc \beta \gamma$ systems, where the actual $\mathcal{N} = 4$ VOA is obtained by additionally taking the kernel of a screening charge \cite{Adamovic:2014lra,Bonetti:2018fqz}. The relation between the two approaches deserves further investigation, where the BRST reduction is split into a two-step process, perhaps by a clever split of the BRST charge \cite{Wolfger:202X}. Furthermore, the reducible module $V^G_{bc\beta \gamma}$ can be projected down to the irreducible submodule $V^G_{\mathcal{N} = 4}$ by an operator $\mathbf{P}$. As a result, the original Schur index
\begin{align}
  \mathcal{I}
  = \operatorname{ch}(V^G_{\mathcal{N} = 4}) = \operatorname{str}_{V^G_{bc\beta \gamma}} \mathbf{P}q^{L_0 - \frac{c_\text{2d}}{24}} b^f \ .
\end{align}
It would be interesting to identify $\mathbf{P}$ and clarify its relation with the screening charge \cite{Bonetti:2018fqz,Adamovic:2014lra} whose kernel gives precisely $V^G_{\mathcal{N} = 4}$. We conjecture that $\mathbf{P}$ can be equivalently replaced by some simple difference (or differential) operator $\mathcal{P}$ acting on the vacuum character $\operatorname{ch}(V^G_{bc\beta \gamma})$, such that
\begin{align}
  \mathcal{I}  = \mathcal{P} \operatorname{str}_{V^G_{bc\beta \gamma}}q^{L_0 - \frac{c_\text{2d}}{24}} b^f \ .
\end{align}
As shown in this paper, the flavored vacuum character on the right coincides with the residue of the integrand of the contour integral that computes $\mathcal{I}$ itself. This conjecture is then equivalent to extracting the full contour integral from the residue of its integrand. This could be achieved by careful application of the elliptic function theory, and as a byproduct, it provides closed-form expressions for the flavored Schur indices. This subject will be studied in \cite{YP:202X}.



\vspace{2em}

\begin{acknowledgments}

We thank Wolfger Peelaers for sharing inspiring observations and discussions. We also thank Yang Lei, Yongchao L\"{u} and Cheng Peng for helpful discussions and comments. Y.P. is supported by the National Natural Science Foundation of China (NSFC) under Grant No. 11905301, the Fundamental Research Funds for the Central Universities under Grant No. 74130-31610023, and the Sun Yat-Sen University Science Foundation.
\end{acknowledgments}

\begin{appendix}
\section{Special functions}

The standard Eisenstein series are defined for $k \in \mathbb{N}_{\ge 1}$ as a $q$-series (where $q = e^{2\pi i \tau}$ throughout this paper)
\begin{align}
  E_{2k} = & \ - \frac{B_{2k}}{(2k)!} + \frac{2}{(2k - 1)!} \sum_{r \ge 1} \frac{r^{2k - 1}q^r}{1 - q^r} \ ,
\end{align}
with $E_\text{odd} = 0$. They are series of integer powers of $q$, and therefore can arise in the modular differential equations in $\mathbb{Z}$ or $\frac{1}{2}\mathbb{Z}$-graded VOAs.

The twisted Eisenstein series are defined for $k \in \mathbb{N}_{\ge 1}$,
\begin{align}
  E_k\left[\begin{matrix}
    \phi \\ \theta
  \end{matrix}\right] = & \ - \frac{B_k(\lambda)}{k!} + \frac{1}{(k - 1)!}\sum'_{r \ge 0} \frac{(r + \lambda)^{k - 1} \theta^{-1} q^{r + \lambda}}{1 - \theta^{-1}q^{r + \lambda}}\nonumber\\
  & \ + \frac{(-1)^k}{(k-1)!} \sum_{r \ge 1} \frac{(r - \lambda)^{k - 1}\theta q^{r - \lambda}}{1 - \theta q^{r - \lambda}} \ .
\end{align}
where $e^{2\pi i\lambda}=\phi$, the prime in the first sum ignores the $r = 0$ term when $\phi = \theta = 1$. They enjoy the symmetry property
\begin{align}
  E_n \left[\begin{matrix}
    \pm 1 \\ \theta^{-1}
  \end{matrix}\right] = (-1)^n E_n \left[\begin{matrix}
    \pm 1 \\ \theta
  \end{matrix}\right] \Rightarrow
  E_\text{odd}\left[\begin{matrix}
    \pm 1\\1
  \end{matrix}\right] = 0 \ .
\end{align}
They are needed for modular differential equations in $\mathbb{R}$-graded VOAs. When $\phi = \theta = 1$, the twisted Eisenstein series with $k = 2n$ simply reduce to $E_{2n}$.

The modular differential operators $D_q^{(k \ge 1)}$ are defined as $\partial_{(2k - 2)}\ldots \partial_{(2)} \partial_{(0)}$ where $\partial_{(k)} \equiv q \partial_q + k E_2$ are the Serre derivatives that map modular forms to higher weight modular forms.

The Jacobi theta functions are defined, in terms of the $q$-Pochhammer symbol $(x;q) \equiv \prod_{k = 0}^{+\infty}(1 - x q^k)$, by
\begin{align}
   \vartheta_1(z) \equiv & \ - i e^{\pi i z}q^{\frac{1}{8}} (q;q)(e^{2\pi i z}q;q)(e^{-2\pi i z};q)\label{def:theta1}\\
   \vartheta_4(z) \equiv & \ (q;q)(e^{2\pi i z}q^{\frac{1}{2}};q)(e^{-2\pi i z} q^{\frac{1}{2}};q)\label{def:theta4}\ .
\end{align}

Here we collect some formula relating $\vartheta$-functions with the Eisenstein series that are useful in massaging the integral identity $\langle T - T_\text{Sug}\rangle = 0$ into a modular differential equation.
\begin{align}
  \frac{\vartheta_1'''(0)}{\vartheta'_1(0)} = & \ 12\pi^2 E_2, \quad
  \frac{\vartheta_4'(\mathfrak{b})}{\vartheta_4(\mathfrak{b})} = 2\pi i E_1\left[
  \begin{matrix}
    -1 \\ b
  \end{matrix}
  \right]\ , \nonumber\\
  \frac{\vartheta_4''(\mathfrak{b})}{\vartheta_4(\mathfrak{b})} = & \ 4\pi^2 \left(E_2 + 2 E_2\left[\begin{matrix}
      -1 \\ b
    \end{matrix}\right]\right) \ , \\
  \frac{\vartheta_1'(\mathfrak{b})}{\vartheta_1(\mathfrak{b})} = & \ 2\pi i E_1 \left[\begin{matrix}
    1 \\ b
  \end{matrix}\right] \ . \nonumber
\end{align}

\section{Flavored modular differential equations\label{app:FMDE}}

A VOA $\mathcal{V}$ is characterized by a space of states $V$ (the \emph{vacuum module}) and a state-operator correspondence $Y$ that builds a local field $Y(a, z)$ out of any state $a \in V$ \cite{Zhu:1996}. We will simply denote the field as $a(z) = \sum_{n \in \mathbb{Z} - h_a} a_n z^{- n - h_a}$ for a weight-$h_a$ state. We also assume the existence and uniqueness of a \emph{vacuum state} $\mathbf{1} \in V$, such that $Y(\mathbf{1}, z) = \operatorname{id}_V$ and $a(0) \mathbf{1} = a$. For a state $a$ with integer weight $h_a$, one defines its zero mode $o(a) = a_0$, whereas $o(a) = 0$ for non-integral $h_a$.

To compute torus correlation functions, it is a common practice to consider $a[z] \equiv e^{i z h_a}Y(a, e^{i z} - 1) = \sum_{n}a_{[n]}z^{-n - h_a}$ where the ``square modes'' $a_{[n]}$ are defined. Explicitly,
\begin{align}
  a_{[n]} = \sum_{j \ge n} c(j, n, h_a)a_j
\end{align}
where the coefficients $c$ are defined by the series expansion
\begin{align}
  (1 + z)^{h - 1}[\log(1 + z)]^n = \sum_{j \ge n} c(j, n, h)z^j\ .
\end{align}
It is worth noting that $o(a_{[-h_a - n]}) = 0$, $\forall n \in \mathbb{N}_{\ge 1}$.

Recursion relations for unflavored torus correlation functions were first studied in \cite{Zhu:1996}, and later generalized to $\mathbb{R}$-graded super-VOAs \cite{Mason:2008zzb} and flavored correlation functions \cite{Gaberdiel:2009vs}. They are the crucial tools for deriving flavored modular differential equations. Consider a $\frac{1}{2}\mathbb{Z}$-graded super-VOA $\mathbf{V}$ containing a $\widehat{\mathfrak{u}}(1)$ current $h$ with zero mode $h_0$, $M$ a module of $\mathbf{V}$ and $a, b \in \mathbf{V}$ are two states of weights $h_a, h_b$. If $h_0 a = 0$, then \footnote{Here all modes are the ``square modes'', which are suitable for torus correlation functions.}\cite{Beem:2017ooy}
\begin{align}
  \label{recursion1}
  & \operatorname{str}_M o(a_{[- h_a]}b)x^{h_0}q^{L_0}
  = \operatorname{str}_M o(a_{[-h_a]} \mathbf{1})o(b) x^{J_0} q^{L_0} \nonumber\\
  & \ \ + \sum_{n = 1}^{+\infty} E_{2k}\left[\begin{matrix}
    e^{2\pi i h_a} \\ 1
  \end{matrix}\right]
  \operatorname{str}_M o(a_{[-h_a + 2k]}b)x^{h_0}q^{L_0} \ .
\end{align}
Recall that when $a$ is a conformal descendant, $o(a_{[-h_a]}) = 0$. The first term plays crucial role when dimensionally reducing torus correlators to topological ones on a circle \cite{Dedushenko:2019mzv,Pan:2019shz}.

If $a$ is charged with $h_0 a = Qa$, then \cite{Mason:2008zzb,Beem:202X}
\begin{align}
  \label{recursion2}
  \operatorname{str}_M & \ o(a_{[- h_a]}b)x^{h_0}q^{L_0} \nonumber \\
  = & \ \sum_{n = 1}^{+\infty} E_n\left[ \begin{matrix}
    e^{2\pi i h_a} \\ x^Q
  \end{matrix} \right]
  \operatorname{str}_M o(a_{[- h_a+n]}b)x^{h_0}q^{L_0}\ .
\end{align}

Using these recursion relations, it is straightforward to write down the modular differential equations associated to the Sugawara relation \cite{Beem:202X}. Suppose that $\mathbf{V}$ contains an affine subalgebra $\widehat{\mathfrak{g}}_k$ and $x^h \equiv x^{\lambda_I H^I} \equiv \prod_{I = 1}^r x_I^{H^I}$. Recalling the standard commutation relations
\begin{align}
  [J^a_{[m]},J^b_{[n]}] = if^{ab}{_c} J^c_{[m + n]} + k m K^{ab} \delta_{m + n, 0} \ ,
\end{align}
where $K$ denotes the Killing form, we have
\begin{align}
  & \ \operatorname{str}_M K_{ab}o(J^a_{[-1]}J^b_{[-1]}\mathbf{1})x^h q^{L_0} \nonumber\\
  = & \ \frac{r}{K(h,h)}\frac{d^2}{dx^2}\operatorname{str}_M x^h q^{L_0} \\
  & \ + (\dim \mathfrak{g} - r) \sum_{\alpha > 0} E_1\left[\begin{matrix}
    1\\ x^{\alpha(h)}
  \end{matrix}\right] K_{IJ}\alpha^I x_J \frac{\partial}{\partial x_J} \operatorname{str}_M x^h q^{L_0} \nonumber\\
  & \ + k(\dim \mathfrak{g} - r)\sum_{\alpha > 0} E_1\left[\begin{matrix}
    1\\ x^{\alpha(h)}
  \end{matrix}\right]\operatorname{str}_M x^h q^{L_0} \nonumber \\
  & \ + k r E_{2} \operatorname{str}_M x^h q^{L_0} \nonumber \ ,
\end{align}
where $\alpha > 0$ denotes all positive roots of $\mathfrak{g}$. For a VOA with a Sugawara relation, we recale the above by the standard factor $\frac{1}{2(k + h^\vee)}$ and equate it with the torus one-point function $ \langle T\rangle = q\partial_q \operatorname{str}_M x^h q^{L_0}$ of the stress tensor, establishing the flavored differential equation. In particular, when $\mathfrak{g} = \mathfrak{su}(2)$, the above reproduces (\ref{fMDE1}) with the variable substitutions $x \to b$, $\operatorname{str}_M x^h q^{L_0 - \frac{c_\text{2d}}{24}} \to I$.

\section{Green's functions}

The action (\ref{twisted-action}) of the torus path integral (\ref{twisted-path-integral}) with non-zero $\mathfrak{a}_\varphi$ leads to twisted Green's function for the $bc$ and $\beta \gamma$ ghosts, given by
\begin{align}
  \label{beta-gamma-greens}
  G_{\beta \gamma}(z)^{\rho \rho'} = & \ \frac{K^{\rho \rho'}e^{i\rho(\mathfrak{a} - \tau \mathfrak{a}_\varphi)\frac{z - \bar z}{\tau - \bar \tau}}}{\pi} \frac{\vartheta'_1(0|\tau)}{\vartheta_4(\rho(\mathfrak{a}-\tau \mathfrak{a}_\varphi) + \mathfrak{b}|\tau)} \nonumber \\ 
  & \ \qquad  \times \frac{\vartheta_4(\frac{z}{2\pi} - \rho(\mathfrak{a} - \tau \mathfrak{a}_\varphi) - \mathfrak{b}|\tau)}{\vartheta_1(\frac{z}{2\pi} | \tau)} \ ,
\end{align}
and
\begin{align}
  \label{bc-greens}
  G_{bc}(z)^{\rho \rho'} = - K^{\rho \rho'} \eta(\tau)^3 & \ e^{ i \rho(\mathfrak{a} - \tau \mathfrak{a}_\varphi) \frac{z - \bar z}{\tau - \bar \tau}}\\
  & \ \times \frac{\vartheta_1(\frac{z}{2\pi} + \rho(\mathfrak{a} - \tau \mathfrak{a}_\varphi) | \tau)}{\vartheta_1(\frac{z}{2\pi}|\tau)\vartheta_1(\rho(\mathfrak{a} - \tau \mathfrak{a}_\varphi | \tau))} \ , \nonumber
\end{align}
where $\rho, \rho' = 0, \pm 2$ labels the the generators of $\mathfrak{su}(2)$.

\end{appendix}

\bibliography{reference}

\begin{thebibliography}{50}%
\makeatletter
\providecommand \@ifxundefined [1]{%
 \@ifx{#1\undefined}
}%
\providecommand \@ifnum [1]{%
 \ifnum #1\expandafter \@firstoftwo
 \else \expandafter \@secondoftwo
 \fi
}%
\providecommand \@ifx [1]{%
 \ifx #1\expandafter \@firstoftwo
 \else \expandafter \@secondoftwo
 \fi
}%
\providecommand \natexlab [1]{#1}%
\providecommand \enquote  [1]{``#1''}%
\providecommand \bibnamefont  [1]{#1}%
\providecommand \bibfnamefont [1]{#1}%
\providecommand \citenamefont [1]{#1}%
\providecommand \href@noop [0]{\@secondoftwo}%
\providecommand \href [0]{\begingroup \@sanitize@url \@href}%
\providecommand \@href[1]{\@@startlink{#1}\@@href}%
\providecommand \@@href[1]{\endgroup#1\@@endlink}%
\providecommand \@sanitize@url [0]{\catcode `\\12\catcode `\$12\catcode
  `\&12\catcode `\#12\catcode `\^12\catcode `\_12\catcode `\%12\relax}%
\providecommand \@@startlink[1]{}%
\providecommand \@@endlink[0]{}%
\providecommand \url  [0]{\begingroup\@sanitize@url \@url }%
\providecommand \@url [1]{\endgroup\@href {#1}{\urlprefix }}%
\providecommand \urlprefix  [0]{URL }%
\providecommand \Eprint [0]{\href }%
\providecommand \doibase [0]{http://dx.doi.org/}%
\providecommand \selectlanguage [0]{\@gobble}%
\providecommand \bibinfo  [0]{\@secondoftwo}%
\providecommand \bibfield  [0]{\@secondoftwo}%
\providecommand \translation [1]{[#1]}%
\providecommand \BibitemOpen [0]{}%
\providecommand \bibitemStop [0]{}%
\providecommand \bibitemNoStop [0]{.\EOS\space}%
\providecommand \EOS [0]{\spacefactor3000\relax}%
\providecommand \BibitemShut  [1]{\csname bibitem#1\endcsname}%
\let\auto@bib@innerbib\@empty
\bibitem [{\citenamefont {Bonetti}\ \emph {et~al.}(2019)\citenamefont
  {Bonetti}, \citenamefont {Meneghelli},\ and\ \citenamefont
  {Rastelli}}]{Bonetti:2018fqz}%
  \BibitemOpen
  \bibfield  {author} {\bibinfo {author} {\bibfnamefont {F.}~\bibnamefont
  {Bonetti}}, \bibinfo {author} {\bibfnamefont {C.}~\bibnamefont {Meneghelli}},
  \ and\ \bibinfo {author} {\bibfnamefont {L.}~\bibnamefont {Rastelli}},\
  }\href {\doibase 10.1007/JHEP05(2019)155} {\bibfield  {journal} {\bibinfo
  {journal} {JHEP}\ }\textbf {\bibinfo {volume} {05}},\ \bibinfo {pages} {155}
  (\bibinfo {year} {2019})},\ \Eprint {http://arxiv.org/abs/1810.03612}
  {arXiv:1810.03612 [hep-th]} \BibitemShut {NoStop}%
\bibitem [{\citenamefont {Beem}\ \emph
  {et~al.}(2015{\natexlab{a}})\citenamefont {Beem}, \citenamefont {Lemos},
  \citenamefont {Liendo}, \citenamefont {Peelaers}, \citenamefont {Rastelli},\
  and\ \citenamefont {van Rees}}]{Beem:2013sza}%
  \BibitemOpen
  \bibfield  {author} {\bibinfo {author} {\bibfnamefont {C.}~\bibnamefont
  {Beem}}, \bibinfo {author} {\bibfnamefont {M.}~\bibnamefont {Lemos}},
  \bibinfo {author} {\bibfnamefont {P.}~\bibnamefont {Liendo}}, \bibinfo
  {author} {\bibfnamefont {W.}~\bibnamefont {Peelaers}}, \bibinfo {author}
  {\bibfnamefont {L.}~\bibnamefont {Rastelli}}, \ and\ \bibinfo {author}
  {\bibfnamefont {B.~C.}\ \bibnamefont {van Rees}},\ }\href {\doibase
  10.1007/s00220-014-2272-x} {\bibfield  {journal} {\bibinfo  {journal}
  {Commun. Math. Phys.}\ }\textbf {\bibinfo {volume} {336}},\ \bibinfo {pages}
  {1359} (\bibinfo {year} {2015}{\natexlab{a}})},\ \Eprint
  {http://arxiv.org/abs/1312.5344} {arXiv:1312.5344 [hep-th]} \BibitemShut
  {NoStop}%
\bibitem [{\citenamefont {Lemos}\ and\ \citenamefont
  {Liendo}(2016)}]{Lemos:2015orc}%
  \BibitemOpen
  \bibfield  {author} {\bibinfo {author} {\bibfnamefont {M.}~\bibnamefont
  {Lemos}}\ and\ \bibinfo {author} {\bibfnamefont {P.}~\bibnamefont {Liendo}},\
  }\href {\doibase 10.1007/JHEP04(2016)004} {\bibfield  {journal} {\bibinfo
  {journal} {JHEP}\ }\textbf {\bibinfo {volume} {04}},\ \bibinfo {pages} {004}
  (\bibinfo {year} {2016})},\ \Eprint {http://arxiv.org/abs/1511.07449}
  {arXiv:1511.07449 [hep-th]} \BibitemShut {NoStop}%
\bibitem [{\citenamefont {Cordova}\ and\ \citenamefont
  {Shao}(2016)}]{Cordova:2015nma}%
  \BibitemOpen
  \bibfield  {author} {\bibinfo {author} {\bibfnamefont {C.}~\bibnamefont
  {Cordova}}\ and\ \bibinfo {author} {\bibfnamefont {S.-H.}\ \bibnamefont
  {Shao}},\ }\href {\doibase 10.1007/JHEP01(2016)040} {\bibfield  {journal}
  {\bibinfo  {journal} {JHEP}\ }\textbf {\bibinfo {volume} {01}},\ \bibinfo
  {pages} {040} (\bibinfo {year} {2016})},\ \Eprint
  {http://arxiv.org/abs/1506.00265} {arXiv:1506.00265 [hep-th]} \BibitemShut
  {NoStop}%
\bibitem [{\citenamefont {Beem}\ \emph
  {et~al.}(2015{\natexlab{b}})\citenamefont {Beem}, \citenamefont {Peelaers},
  \citenamefont {Rastelli},\ and\ \citenamefont {van Rees}}]{Beem:2014rza}%
  \BibitemOpen
  \bibfield  {author} {\bibinfo {author} {\bibfnamefont {C.}~\bibnamefont
  {Beem}}, \bibinfo {author} {\bibfnamefont {W.}~\bibnamefont {Peelaers}},
  \bibinfo {author} {\bibfnamefont {L.}~\bibnamefont {Rastelli}}, \ and\
  \bibinfo {author} {\bibfnamefont {B.~C.}\ \bibnamefont {van Rees}},\ }\href
  {\doibase 10.1007/JHEP05(2015)020} {\bibfield  {journal} {\bibinfo  {journal}
  {JHEP}\ }\textbf {\bibinfo {volume} {05}},\ \bibinfo {pages} {020} (\bibinfo
  {year} {2015}{\natexlab{b}})},\ \Eprint {http://arxiv.org/abs/1408.6522}
  {arXiv:1408.6522 [hep-th]} \BibitemShut {NoStop}%
\bibitem [{\citenamefont {Lemos}\ and\ \citenamefont
  {Peelaers}(2015)}]{Lemos:2014lua}%
  \BibitemOpen
  \bibfield  {author} {\bibinfo {author} {\bibfnamefont {M.}~\bibnamefont
  {Lemos}}\ and\ \bibinfo {author} {\bibfnamefont {W.}~\bibnamefont
  {Peelaers}},\ }\href {\doibase 10.1007/JHEP02(2015)113} {\bibfield  {journal}
  {\bibinfo  {journal} {JHEP}\ }\textbf {\bibinfo {volume} {02}},\ \bibinfo
  {pages} {113} (\bibinfo {year} {2015})},\ \Eprint
  {http://arxiv.org/abs/1411.3252} {arXiv:1411.3252 [hep-th]} \BibitemShut
  {NoStop}%
\bibitem [{\citenamefont {Song}\ \emph {et~al.}(2017)\citenamefont {Song},
  \citenamefont {Xie},\ and\ \citenamefont {Yan}}]{Song:2017oew}%
  \BibitemOpen
  \bibfield  {author} {\bibinfo {author} {\bibfnamefont {J.}~\bibnamefont
  {Song}}, \bibinfo {author} {\bibfnamefont {D.}~\bibnamefont {Xie}}, \ and\
  \bibinfo {author} {\bibfnamefont {W.}~\bibnamefont {Yan}},\ }\href {\doibase
  10.1007/JHEP12(2017)123} {\bibfield  {journal} {\bibinfo  {journal} {JHEP}\
  }\textbf {\bibinfo {volume} {12}},\ \bibinfo {pages} {123} (\bibinfo {year}
  {2017})},\ \Eprint {http://arxiv.org/abs/1706.01607} {arXiv:1706.01607
  [hep-th]} \BibitemShut {NoStop}%
\bibitem [{\citenamefont {Xie}\ and\ \citenamefont
  {Yan}(2019{\natexlab{a}})}]{Xie:2019yds}%
  \BibitemOpen
  \bibfield  {author} {\bibinfo {author} {\bibfnamefont {D.}~\bibnamefont
  {Xie}}\ and\ \bibinfo {author} {\bibfnamefont {W.}~\bibnamefont {Yan}},\
  }\href@noop {} {\  (\bibinfo {year} {2019}{\natexlab{a}})},\ \Eprint
  {http://arxiv.org/abs/1902.02838} {arXiv:1902.02838 [hep-th]} \BibitemShut
  {NoStop}%
\bibitem [{\citenamefont {Xie}\ and\ \citenamefont
  {Yan}(2019{\natexlab{b}})}]{Xie:2019vzr}%
  \BibitemOpen
  \bibfield  {author} {\bibinfo {author} {\bibfnamefont {D.}~\bibnamefont
  {Xie}}\ and\ \bibinfo {author} {\bibfnamefont {W.}~\bibnamefont {Yan}},\
  }\href@noop {} {\  (\bibinfo {year} {2019}{\natexlab{b}})},\ \Eprint
  {http://arxiv.org/abs/1910.02281} {arXiv:1910.02281 [hep-th]} \BibitemShut
  {NoStop}%
\bibitem [{\citenamefont {Creutzig}(2017)}]{Creutzig:2017qyf}%
  \BibitemOpen
  \bibfield  {author} {\bibinfo {author} {\bibfnamefont {T.}~\bibnamefont
  {Creutzig}},\ }\href@noop {} {\  (\bibinfo {year} {2017})},\ \Eprint
  {http://arxiv.org/abs/1701.05926} {arXiv:1701.05926 [hep-th]} \BibitemShut
  {NoStop}%
\bibitem [{\citenamefont {Beem}\ \emph {et~al.}(2019)\citenamefont {Beem},
  \citenamefont {Meneghelli},\ and\ \citenamefont {Rastelli}}]{Beem:2019tfp}%
  \BibitemOpen
  \bibfield  {author} {\bibinfo {author} {\bibfnamefont {C.}~\bibnamefont
  {Beem}}, \bibinfo {author} {\bibfnamefont {C.}~\bibnamefont {Meneghelli}}, \
  and\ \bibinfo {author} {\bibfnamefont {L.}~\bibnamefont {Rastelli}},\ }\href
  {\doibase 10.1007/JHEP09(2019)058} {\bibfield  {journal} {\bibinfo  {journal}
  {JHEP}\ }\textbf {\bibinfo {volume} {09}},\ \bibinfo {pages} {058} (\bibinfo
  {year} {2019})},\ \Eprint {http://arxiv.org/abs/1903.07624} {arXiv:1903.07624
  [hep-th]} \BibitemShut {NoStop}%
\bibitem [{\citenamefont {Beem}\ \emph {et~al.}(2020)\citenamefont {Beem},
  \citenamefont {Meneghelli}, \citenamefont {Peelaers},\ and\ \citenamefont
  {Rastelli}}]{Beem:2019snk}%
  \BibitemOpen
  \bibfield  {author} {\bibinfo {author} {\bibfnamefont {C.}~\bibnamefont
  {Beem}}, \bibinfo {author} {\bibfnamefont {C.}~\bibnamefont {Meneghelli}},
  \bibinfo {author} {\bibfnamefont {W.}~\bibnamefont {Peelaers}}, \ and\
  \bibinfo {author} {\bibfnamefont {L.}~\bibnamefont {Rastelli}},\ }\href
  {\doibase 10.1007/s00220-020-03746-9} {\bibfield  {journal} {\bibinfo
  {journal} {Commun. Math. Phys.}\ }\textbf {\bibinfo {volume} {377}},\
  \bibinfo {pages} {2553} (\bibinfo {year} {2020})},\ \Eprint
  {http://arxiv.org/abs/1907.08629} {arXiv:1907.08629 [hep-th]} \BibitemShut
  {NoStop}%
\bibitem [{\citenamefont {Cordova}\ \emph {et~al.}(2017)\citenamefont
  {Cordova}, \citenamefont {Gaiotto},\ and\ \citenamefont
  {Shao}}]{Cordova:2017mhb}%
  \BibitemOpen
  \bibfield  {author} {\bibinfo {author} {\bibfnamefont {C.}~\bibnamefont
  {Cordova}}, \bibinfo {author} {\bibfnamefont {D.}~\bibnamefont {Gaiotto}}, \
  and\ \bibinfo {author} {\bibfnamefont {S.-H.}\ \bibnamefont {Shao}},\ }\href
  {\doibase 10.1007/JHEP05(2017)140} {\bibfield  {journal} {\bibinfo  {journal}
  {JHEP}\ }\textbf {\bibinfo {volume} {05}},\ \bibinfo {pages} {140} (\bibinfo
  {year} {2017})},\ \Eprint {http://arxiv.org/abs/1704.01955} {arXiv:1704.01955
  [hep-th]} \BibitemShut {NoStop}%
\bibitem [{\citenamefont {Beem}\ \emph {et~al.}()\citenamefont {Beem},
  \citenamefont {Peelaers},\ and\ \citenamefont {Rastelli}}]{Beem:203X}%
  \BibitemOpen
  \bibfield  {author} {\bibinfo {author} {\bibfnamefont {C.}~\bibnamefont
  {Beem}}, \bibinfo {author} {\bibfnamefont {W.}~\bibnamefont {Peelaers}}, \
  and\ \bibinfo {author} {\bibfnamefont {L.}~\bibnamefont {Rastelli}},\
  }\href@noop {} {\ }\Eprint {http://arxiv.org/abs/unpublished} {unpublished}
  \BibitemShut {NoStop}%
\bibitem [{\citenamefont {Cordova}\ \emph {et~al.}(2016)\citenamefont
  {Cordova}, \citenamefont {Gaiotto},\ and\ \citenamefont
  {Shao}}]{Cordova:2016uwk}%
  \BibitemOpen
  \bibfield  {author} {\bibinfo {author} {\bibfnamefont {C.}~\bibnamefont
  {Cordova}}, \bibinfo {author} {\bibfnamefont {D.}~\bibnamefont {Gaiotto}}, \
  and\ \bibinfo {author} {\bibfnamefont {S.-H.}\ \bibnamefont {Shao}},\ }\href
  {\doibase 10.1007/JHEP11(2016)106} {\bibfield  {journal} {\bibinfo  {journal}
  {JHEP}\ }\textbf {\bibinfo {volume} {11}},\ \bibinfo {pages} {106} (\bibinfo
  {year} {2016})},\ \Eprint {http://arxiv.org/abs/1606.08429} {arXiv:1606.08429
  [hep-th]} \BibitemShut {NoStop}%
\bibitem [{\citenamefont {Nishinaka}\ \emph {et~al.}(2019)\citenamefont
  {Nishinaka}, \citenamefont {Sasa},\ and\ \citenamefont
  {Zhu}}]{Nishinaka:2018zwq}%
  \BibitemOpen
  \bibfield  {author} {\bibinfo {author} {\bibfnamefont {T.}~\bibnamefont
  {Nishinaka}}, \bibinfo {author} {\bibfnamefont {S.}~\bibnamefont {Sasa}}, \
  and\ \bibinfo {author} {\bibfnamefont {R.-D.}\ \bibnamefont {Zhu}},\ }\href
  {\doibase 10.1007/JHEP03(2019)091} {\bibfield  {journal} {\bibinfo  {journal}
  {JHEP}\ }\textbf {\bibinfo {volume} {03}},\ \bibinfo {pages} {091} (\bibinfo
  {year} {2019})},\ \Eprint {http://arxiv.org/abs/1811.11772} {arXiv:1811.11772
  [hep-th]} \BibitemShut {NoStop}%
\bibitem [{\citenamefont {Pan}\ and\ \citenamefont
  {Peelaers}(2018)}]{Pan:2017zie}%
  \BibitemOpen
  \bibfield  {author} {\bibinfo {author} {\bibfnamefont {Y.}~\bibnamefont
  {Pan}}\ and\ \bibinfo {author} {\bibfnamefont {W.}~\bibnamefont {Peelaers}},\
  }\href {\doibase 10.1007/JHEP02(2018)138} {\bibfield  {journal} {\bibinfo
  {journal} {JHEP}\ }\textbf {\bibinfo {volume} {02}},\ \bibinfo {pages} {138}
  (\bibinfo {year} {2018})},\ \Eprint {http://arxiv.org/abs/1710.04306}
  {arXiv:1710.04306 [hep-th]} \BibitemShut {NoStop}%
\bibitem [{\citenamefont {Dedushenko}\ and\ \citenamefont
  {Fluder}(2020)}]{Dedushenko:2019yiw}%
  \BibitemOpen
  \bibfield  {author} {\bibinfo {author} {\bibfnamefont {M.}~\bibnamefont
  {Dedushenko}}\ and\ \bibinfo {author} {\bibfnamefont {M.}~\bibnamefont
  {Fluder}},\ }\href {\doibase 10.1063/5.0002661} {\bibfield  {journal}
  {\bibinfo  {journal} {J. Math. Phys.}\ }\textbf {\bibinfo {volume} {61}},\
  \bibinfo {pages} {092302} (\bibinfo {year} {2020})},\ \Eprint
  {http://arxiv.org/abs/1904.02704} {arXiv:1904.02704 [hep-th]} \BibitemShut
  {NoStop}%
\bibitem [{\citenamefont {Gaiotto}\ \emph {et~al.}(2013)\citenamefont
  {Gaiotto}, \citenamefont {Rastelli},\ and\ \citenamefont
  {Razamat}}]{Gaiotto:2012xa}%
  \BibitemOpen
  \bibfield  {author} {\bibinfo {author} {\bibfnamefont {D.}~\bibnamefont
  {Gaiotto}}, \bibinfo {author} {\bibfnamefont {L.}~\bibnamefont {Rastelli}}, \
  and\ \bibinfo {author} {\bibfnamefont {S.~S.}\ \bibnamefont {Razamat}},\
  }\href {\doibase 10.1007/JHEP01(2013)022} {\bibfield  {journal} {\bibinfo
  {journal} {JHEP}\ }\textbf {\bibinfo {volume} {01}},\ \bibinfo {pages} {022}
  (\bibinfo {year} {2013})},\ \Eprint {http://arxiv.org/abs/1207.3577}
  {arXiv:1207.3577 [hep-th]} \BibitemShut {NoStop}%
\bibitem [{\citenamefont {Gukov}\ and\ \citenamefont
  {Witten}(2010)}]{Gukov:2008sn}%
  \BibitemOpen
  \bibfield  {author} {\bibinfo {author} {\bibfnamefont {S.}~\bibnamefont
  {Gukov}}\ and\ \bibinfo {author} {\bibfnamefont {E.}~\bibnamefont {Witten}},\
  }\href {\doibase 10.4310/ATMP.2010.v14.n1.a3} {\bibfield  {journal} {\bibinfo
   {journal} {Adv. Theor. Math. Phys.}\ }\textbf {\bibinfo {volume} {14}},\
  \bibinfo {pages} {87} (\bibinfo {year} {2010})},\ \Eprint
  {http://arxiv.org/abs/0804.1561} {arXiv:0804.1561 [hep-th]} \BibitemShut
  {NoStop}%
\bibitem [{\citenamefont {Drukker}\ \emph {et~al.}(2014)\citenamefont
  {Drukker}, \citenamefont {Okuda},\ and\ \citenamefont
  {Passerini}}]{Drukker:2012sr}%
  \BibitemOpen
  \bibfield  {author} {\bibinfo {author} {\bibfnamefont {N.}~\bibnamefont
  {Drukker}}, \bibinfo {author} {\bibfnamefont {T.}~\bibnamefont {Okuda}}, \
  and\ \bibinfo {author} {\bibfnamefont {F.}~\bibnamefont {Passerini}},\ }\href
  {\doibase 10.1007/JHEP07(2014)137} {\bibfield  {journal} {\bibinfo  {journal}
  {JHEP}\ }\textbf {\bibinfo {volume} {07}},\ \bibinfo {pages} {137} (\bibinfo
  {year} {2014})},\ \Eprint {http://arxiv.org/abs/1211.3409} {arXiv:1211.3409
  [hep-th]} \BibitemShut {NoStop}%
\bibitem [{\citenamefont {Kapustin}\ \emph {et~al.}(2013)\citenamefont
  {Kapustin}, \citenamefont {Willett},\ and\ \citenamefont
  {Yaakov}}]{Kapustin:2012iw}%
  \BibitemOpen
  \bibfield  {author} {\bibinfo {author} {\bibfnamefont {A.}~\bibnamefont
  {Kapustin}}, \bibinfo {author} {\bibfnamefont {B.}~\bibnamefont {Willett}}, \
  and\ \bibinfo {author} {\bibfnamefont {I.}~\bibnamefont {Yaakov}},\ }\href
  {\doibase 10.1007/JHEP06(2013)099} {\bibfield  {journal} {\bibinfo  {journal}
  {JHEP}\ }\textbf {\bibinfo {volume} {06}},\ \bibinfo {pages} {099} (\bibinfo
  {year} {2013})},\ \Eprint {http://arxiv.org/abs/1211.2861} {arXiv:1211.2861
  [hep-th]} \BibitemShut {NoStop}%
\bibitem [{\citenamefont {Nawata}(2015)}]{Nawata:2014nca}%
  \BibitemOpen
  \bibfield  {author} {\bibinfo {author} {\bibfnamefont {S.}~\bibnamefont
  {Nawata}},\ }\href {\doibase 10.4310/ATMP.2015.v19.n6.a4} {\bibfield
  {journal} {\bibinfo  {journal} {Adv. Theor. Math. Phys.}\ }\textbf {\bibinfo
  {volume} {19}},\ \bibinfo {pages} {1277} (\bibinfo {year} {2015})},\ \Eprint
  {http://arxiv.org/abs/1408.4132} {arXiv:1408.4132 [hep-th]} \BibitemShut
  {NoStop}%
\bibitem [{\citenamefont {Hosomichi}\ \emph {et~al.}(2018)\citenamefont
  {Hosomichi}, \citenamefont {Lee},\ and\ \citenamefont
  {Okuda}}]{Hosomichi:2017dbc}%
  \BibitemOpen
  \bibfield  {author} {\bibinfo {author} {\bibfnamefont {K.}~\bibnamefont
  {Hosomichi}}, \bibinfo {author} {\bibfnamefont {S.}~\bibnamefont {Lee}}, \
  and\ \bibinfo {author} {\bibfnamefont {T.}~\bibnamefont {Okuda}},\ }\href
  {\doibase 10.1007/JHEP01(2018)033} {\bibfield  {journal} {\bibinfo  {journal}
  {JHEP}\ }\textbf {\bibinfo {volume} {01}},\ \bibinfo {pages} {033} (\bibinfo
  {year} {2018})},\ \Eprint {http://arxiv.org/abs/1705.10623} {arXiv:1705.10623
  [hep-th]} \BibitemShut {NoStop}%
\bibitem [{Note1()}]{Note1}%
  \BibitemOpen
  \bibinfo {note} {W. Peelaers, private communication on the cases with
  $\protect \mathcal {N} = 4$ $SU(2)$ and $SU(2)$ SQCD theories.}\BibitemShut
  {Stop}%
\bibitem [{\citenamefont {Beem}\ and\ \citenamefont
  {Rastelli}(2018)}]{Beem:2017ooy}%
  \BibitemOpen
  \bibfield  {author} {\bibinfo {author} {\bibfnamefont {C.}~\bibnamefont
  {Beem}}\ and\ \bibinfo {author} {\bibfnamefont {L.}~\bibnamefont
  {Rastelli}},\ }\href {\doibase 10.1007/JHEP08(2018)114} {\bibfield  {journal}
  {\bibinfo  {journal} {JHEP}\ }\textbf {\bibinfo {volume} {08}},\ \bibinfo
  {pages} {114} (\bibinfo {year} {2018})},\ \Eprint
  {http://arxiv.org/abs/1707.07679} {arXiv:1707.07679 [hep-th]} \BibitemShut
  {NoStop}%
\bibitem [{\citenamefont {Pan}\ and\ \citenamefont
  {Peelaers}(2019)}]{Pan:2019bor}%
  \BibitemOpen
  \bibfield  {author} {\bibinfo {author} {\bibfnamefont {Y.}~\bibnamefont
  {Pan}}\ and\ \bibinfo {author} {\bibfnamefont {W.}~\bibnamefont {Peelaers}},\
  }\href {\doibase 10.1007/JHEP07(2019)013} {\bibfield  {journal} {\bibinfo
  {journal} {JHEP}\ }\textbf {\bibinfo {volume} {07}},\ \bibinfo {pages} {013}
  (\bibinfo {year} {2019})},\ \Eprint {http://arxiv.org/abs/1903.03623}
  {arXiv:1903.03623 [hep-th]} \BibitemShut {NoStop}%
\bibitem [{Note2()}]{Note2}%
  \BibitemOpen
  \bibinfo {note} {Here the prime in the measure indicates that we drop the $c$
  zero mode when it is $\protect \mathfrak {su}(2)$ Cartan-valued.}\BibitemShut
  {Stop}%
\bibitem [{Note3()}]{Note3}%
  \BibitemOpen
  \bibinfo {note} {When writing the integrand in terms of $q$-Pochhammer
  symbols, some polynomial factors from the $\vartheta _1$'s will join
  $[da/2\pi i a]$ to form the standard Haar measure.}\BibitemShut {Stop}%
\bibitem [{Note4()}]{Note4}%
  \BibitemOpen
  \bibinfo {note} {A complete treatment involves first removing a tubular
  neighborhood of the surface defect and consider suitable fields and
  supersymmetric theory on that space, along the line as \cite
  {Drukker:2012sr,Hosomichi:2017dbc}. The neighborhood does not intersect the
  $T^2$ where the $bc \beta \gamma $ system lives, and we conjecture that the
  only effect left on $T^2$ is the background gauge field $\protect \mathfrak
  {a}_\varphi $. We leave the precise computation to future work.}\BibitemShut
  {Stop}%
\bibitem [{\citenamefont {Eholzer}\ \emph {et~al.}(1998)\citenamefont
  {Eholzer}, \citenamefont {Feher},\ and\ \citenamefont
  {Honecker}}]{Eholzer:1997se}%
  \BibitemOpen
  \bibfield  {author} {\bibinfo {author} {\bibfnamefont {W.}~\bibnamefont
  {Eholzer}}, \bibinfo {author} {\bibfnamefont {L.}~\bibnamefont {Feher}}, \
  and\ \bibinfo {author} {\bibfnamefont {A.}~\bibnamefont {Honecker}},\ }\href
  {\doibase 10.1016/S0550-3213(98)00061-3} {\bibfield  {journal} {\bibinfo
  {journal} {Nucl. Phys. B}\ }\textbf {\bibinfo {volume} {518}},\ \bibinfo
  {pages} {669} (\bibinfo {year} {1998})},\ \Eprint
  {http://arxiv.org/abs/hep-th/9708160} {arXiv:hep-th/9708160} \BibitemShut
  {NoStop}%
\bibitem [{Note5()}]{Note5}%
  \BibitemOpen
  \bibinfo {note} {Note that another potential pole $(a_1 = a_2
  b^{-1}q^{\protect \frac {1}{2}}, a_2 = a_1^{- 2} b^{-1}q^{\protect \frac
  {1}{2}})$ is canceled by the $\vartheta _1(\protect \mathfrak {a}_2 -
  \protect \mathfrak {a}_3)^2$ factor.}\BibitemShut {Stop}%
\bibitem [{Note6()}]{Note6}%
  \BibitemOpen
  \bibinfo {note} {Although equivalent as analytic functions, these residues
  are different as $\protect \mathfrak {b}$-series with specific convergence
  annulus. We hope to come back to this issue in future work.}\BibitemShut
  {Stop}%
\bibitem [{\citenamefont {Goldstein}\ \emph {et~al.}(2020)\citenamefont
  {Goldstein}, \citenamefont {Jejjala}, \citenamefont {Lei}, \citenamefont {van
  Leuven},\ and\ \citenamefont {Li}}]{Goldstein:2020yvj}%
  \BibitemOpen
  \bibfield  {author} {\bibinfo {author} {\bibfnamefont {K.}~\bibnamefont
  {Goldstein}}, \bibinfo {author} {\bibfnamefont {V.}~\bibnamefont {Jejjala}},
  \bibinfo {author} {\bibfnamefont {Y.}~\bibnamefont {Lei}}, \bibinfo {author}
  {\bibfnamefont {S.}~\bibnamefont {van Leuven}}, \ and\ \bibinfo {author}
  {\bibfnamefont {W.}~\bibnamefont {Li}},\ }\href@noop {} {\  (\bibinfo {year}
  {2020})},\ \Eprint {http://arxiv.org/abs/2011.06605} {arXiv:2011.06605
  [hep-th]} \BibitemShut {NoStop}%
\bibitem [{\citenamefont {Collingwood}\ and\ \citenamefont
  {McGovern}()}]{Collingwood}%
  \BibitemOpen
  \bibfield  {author} {\bibinfo {author} {\bibfnamefont {D.}~\bibnamefont
  {Collingwood}}\ and\ \bibinfo {author} {\bibfnamefont {W.}~\bibnamefont
  {McGovern}},\ }\href@noop {} {\ }\Eprint {http://arxiv.org/abs/{Nilpotent
  orbits in semisimple Lie algebras (1993)}} {{Nilpotent orbits in semisimple
  Lie algebras (1993)}} \BibitemShut {NoStop}%
\bibitem [{Note7()}]{Note7}%
  \BibitemOpen
  \bibinfo {note} {Note the special multiplicity of $2$ for the $\protect
  \mathfrak {d}_r$ case at $H = r - 1$, since $|\Delta _{r - 1}^+| - |\Delta
  _r^+| = 2$, so the corresponding factor in the above product should be
  squared.}\BibitemShut {Stop}%
\bibitem [{\citenamefont {Gaberdiel}\ and\ \citenamefont
  {Keller}(2008)}]{Gaberdiel:2008pr}%
  \BibitemOpen
  \bibfield  {author} {\bibinfo {author} {\bibfnamefont {M.~R.}\ \bibnamefont
  {Gaberdiel}}\ and\ \bibinfo {author} {\bibfnamefont {C.~A.}\ \bibnamefont
  {Keller}},\ }\href {\doibase 10.1088/1126-6708/2008/09/079} {\bibfield
  {journal} {\bibinfo  {journal} {JHEP}\ }\textbf {\bibinfo {volume} {09}},\
  \bibinfo {pages} {079} (\bibinfo {year} {2008})},\ \Eprint
  {http://arxiv.org/abs/0804.0489} {arXiv:0804.0489 [hep-th]} \BibitemShut
  {NoStop}%
\bibitem [{\citenamefont {Gaberdiel}\ and\ \citenamefont
  {Lang}(2009)}]{Gaberdiel:2008ma}%
  \BibitemOpen
  \bibfield  {author} {\bibinfo {author} {\bibfnamefont {M.~R.}\ \bibnamefont
  {Gaberdiel}}\ and\ \bibinfo {author} {\bibfnamefont {S.}~\bibnamefont
  {Lang}},\ }\href {\doibase 10.1088/1751-8113/42/4/045405} {\bibfield
  {journal} {\bibinfo  {journal} {J. Phys. A}\ }\textbf {\bibinfo {volume}
  {42}},\ \bibinfo {pages} {045405} (\bibinfo {year} {2009})},\ \Eprint
  {http://arxiv.org/abs/0810.0106} {arXiv:0810.0106 [hep-th]} \BibitemShut
  {NoStop}%
\bibitem [{\citenamefont {Beem}\ and\ \citenamefont {Peelaers}()}]{Beem:202X}%
  \BibitemOpen
  \bibfield  {author} {\bibinfo {author} {\bibfnamefont {C.}~\bibnamefont
  {Beem}}\ and\ \bibinfo {author} {\bibfnamefont {W.}~\bibnamefont
  {Peelaers}},\ }\href@noop {} {\ }\Eprint {http://arxiv.org/abs/work in
  progress} {work in progress} \BibitemShut {NoStop}%
\bibitem [{\citenamefont {Arakawa}\ and\ \citenamefont
  {Kawasetsu}(2016)}]{Arakawa:2016hkg}%
  \BibitemOpen
  \bibfield  {author} {\bibinfo {author} {\bibfnamefont {T.}~\bibnamefont
  {Arakawa}}\ and\ \bibinfo {author} {\bibfnamefont {K.}~\bibnamefont
  {Kawasetsu}},\ }\href@noop {} {\  (\bibinfo {year} {2016})},\ \Eprint
  {http://arxiv.org/abs/1610.05865} {arXiv:1610.05865 [math.QA]} \BibitemShut
  {NoStop}%
\bibitem [{Note8()}]{Note8}%
  \BibitemOpen
  \bibinfo {note} {With help from Mathematica to establish some integral
  identities between integrated $q$-series. It would be great to actually prove
  the equality analytically.}\BibitemShut {Stop}%
\bibitem [{\citenamefont {Adamovic}(2014)}]{Adamovic:2014lra}%
  \BibitemOpen
  \bibfield  {author} {\bibinfo {author} {\bibfnamefont {D.}~\bibnamefont
  {Adamovic}},\ }\href@noop {} {\  (\bibinfo {year} {2014})},\ \Eprint
  {http://arxiv.org/abs/1407.1527} {arXiv:1407.1527 [math.QA]} \BibitemShut
  {NoStop}%
\bibitem [{\citenamefont {Pan}\ and\ \citenamefont {Peelaers}()}]{YP:202X}%
  \BibitemOpen
  \bibfield  {author} {\bibinfo {author} {\bibfnamefont {Y.}~\bibnamefont
  {Pan}}\ and\ \bibinfo {author} {\bibfnamefont {W.}~\bibnamefont {Peelaers}},\
  }\href@noop {} {\ }\Eprint {http://arxiv.org/abs/work in progress} {work in
  progress} \BibitemShut {NoStop}%
\bibitem [{\citenamefont {Peelaers}()}]{Wolfger:202X}%
  \BibitemOpen
  \bibfield  {author} {\bibinfo {author} {\bibfnamefont {W.}~\bibnamefont
  {Peelaers}},\ }\href@noop {} {\ }\Eprint {http://arxiv.org/abs/Private
  communication} {Private communication} \BibitemShut {NoStop}%
\bibitem [{\citenamefont {Zhu}(1996)}]{Zhu:1996}%
  \BibitemOpen
  \bibfield  {author} {\bibinfo {author} {\bibfnamefont {Y.}~\bibnamefont
  {Zhu}},\ }\href {\doibase 10.1090/S0894-0347-96-00182-8} {\bibfield
  {journal} {\bibinfo  {journal} {J. Amer. Math. Soc.}\ }\textbf {\bibinfo
  {volume} {9}},\ \bibinfo {pages} {237} (\bibinfo {year} {1996})}\BibitemShut
  {NoStop}%
\bibitem [{\citenamefont {Mason}\ \emph {et~al.}(2008)\citenamefont {Mason},
  \citenamefont {Tuite},\ and\ \citenamefont {Zuevsky}}]{Mason:2008zzb}%
  \BibitemOpen
  \bibfield  {author} {\bibinfo {author} {\bibfnamefont {G.}~\bibnamefont
  {Mason}}, \bibinfo {author} {\bibfnamefont {M.~P.}\ \bibnamefont {Tuite}}, \
  and\ \bibinfo {author} {\bibfnamefont {A.}~\bibnamefont {Zuevsky}},\ }\href
  {\doibase 10.1007/s00220-008-0510-9} {\bibfield  {journal} {\bibinfo
  {journal} {Commun. Math. Phys.}\ }\textbf {\bibinfo {volume} {283}},\
  \bibinfo {pages} {305} (\bibinfo {year} {2008})},\ \Eprint
  {http://arxiv.org/abs/0708.0640} {arXiv:0708.0640 [math.QA]} \BibitemShut
  {NoStop}%
\bibitem [{\citenamefont {Gaberdiel}\ and\ \citenamefont
  {Keller}(2009)}]{Gaberdiel:2009vs}%
  \BibitemOpen
  \bibfield  {author} {\bibinfo {author} {\bibfnamefont {M.~R.}\ \bibnamefont
  {Gaberdiel}}\ and\ \bibinfo {author} {\bibfnamefont {C.~A.}\ \bibnamefont
  {Keller}},\ }\href {\doibase 10.4310/CNTP.2009.v3.n4.a1} {\bibfield
  {journal} {\bibinfo  {journal} {Commun. Num. Theor. Phys.}\ }\textbf
  {\bibinfo {volume} {3}},\ \bibinfo {pages} {593} (\bibinfo {year} {2009})},\
  \Eprint {http://arxiv.org/abs/0904.1831} {arXiv:0904.1831 [hep-th]}
  \BibitemShut {NoStop}%
\bibitem [{Note9()}]{Note9}%
  \BibitemOpen
  \bibinfo {note} {Here all modes are the ``square modes'', which are suitable
  for torus correlation functions.}\BibitemShut {Stop}%
\bibitem [{\citenamefont {Dedushenko}(2019)}]{Dedushenko:2019mzv}%
  \BibitemOpen
  \bibfield  {author} {\bibinfo {author} {\bibfnamefont {M.}~\bibnamefont
  {Dedushenko}},\ }\href@noop {} {\  (\bibinfo {year} {2019})},\ \Eprint
  {http://arxiv.org/abs/1911.05741} {arXiv:1911.05741 [hep-th]} \BibitemShut
  {NoStop}%
\bibitem [{\citenamefont {Pan}\ and\ \citenamefont
  {Peelaers}(2020)}]{Pan:2019shz}%
  \BibitemOpen
  \bibfield  {author} {\bibinfo {author} {\bibfnamefont {Y.}~\bibnamefont
  {Pan}}\ and\ \bibinfo {author} {\bibfnamefont {W.}~\bibnamefont {Peelaers}},\
  }\href {\doibase 10.1007/JHEP06(2020)127} {\bibfield  {journal} {\bibinfo
  {journal} {JHEP}\ }\textbf {\bibinfo {volume} {06}},\ \bibinfo {pages} {127}
  (\bibinfo {year} {2020})},\ \Eprint {http://arxiv.org/abs/1911.09631}
  {arXiv:1911.09631 [hep-th]} \BibitemShut {NoStop}%
\end{thebibliography}%

\end{document}